\documentclass{article}%
\usepackage{amsmath}
\usepackage{amsfonts}
\usepackage{amssymb}
\usepackage{graphicx}%
\begin{document}

\begin{center}
{\Large The Hamilton-Jacobi Approach to Teleparallelism}\newline%
\newline\bigskip B. M. Pimentel$^{(a)}$\footnote{pimentel@ift.unesp.br}, P. J.
Pompeia$^{(a,c)}$\footnote{pompeia@ift.unesp.br}, J. F. da Rocha-Neto$^{(b)}%
$\footnote{rocha@webmail.fis.unb.br}\newline$^{(a)}$ Instituto de F\'{\i}sica
Te\'{o}rica - Universidade Estadual Paulista,\newline Rua Pamplona 145,
01405-900, S\~{a}o Paulo, SP, Brazil.\newline$^{(b)}$ Campus Universit\'{a}rio
de Arraias - Universidade Federal do Tocantins,\newline Rua Universit\'{a}ria
s/n$^{0}$ Centro, 77.330-000, Arraias, TO, Brazil.\newline$^{(c)}$ Centro
T\'{e}cnico Aeroespacial - Instituto de Fomento e Coordena\c{c}\~{a}o
Industrial - Divis\~{a}o de Confiabilidade Metrol\'{o}gica\newline Pra\c{c}a
Mal. Eduardo Gomes, 50, 12228-901, S\~{a}o Jos\'{e} dos Campos, SP,
Brazil.\newline

\bigskip\textit{Abstract:}
\end{center}

\begin{quotation}
\textit{We intend to analyse the constraint structure of Teleparallelism
employing the Hamilton-Jacobi formalism for singular systems. This study is
conducted without using an ADM 3+1 decomposition and without fixing time gauge
condition. It can be verified that the field equations constitute an
integrable system.}

\bigskip

Keywords: Hamilton-Jacobi; Teleparallelism; Constraints.
\end{quotation}

\section{Introduction}

The standard approach to study singular (constrained) systems is that one
developed by Dirac \cite{Dirac0}, \cite{Dirac 0.5}, \cite{Dirac 0.75} in which
the hamiltonian structure is employed. The success of this approach cannot be
denied if we consider the wide number of cases in which it has been
successfully applied. Among these cases we can cite Podolsky Electrodynamics
\cite{Podolsky}, General Relativity \cite{Dirac1}, \cite{Dirac2} and
Teleparallelism \cite{Maluf & RN}, \cite{Maluf}, \cite{Blagojevic}. One
important feature of this formalism lies in the physical arguments provided by
consistency conditions, which tell that the constraints of the theory must be
conserved, \textit{i.e.}, their time derivatives must be null.\ Although Dirac
approach is widely accepted \cite{Regge}, \cite{Sunder}, \cite{Gitman},
\cite{Govaerts}, \cite{Henneaux}, it did not avoid the emerging of other
approaches, that always provide new points of view for the same problems.

One of these approaches is the Hamilton-Jacobi (HJ) formalism, that makes use
of Carath\'{e}odory Equivalent Lagrangian method \cite{Carat}. This method is
an alternative way to obtain Hamilton-Jacobi equation starting from lagrangian
formalism. It was originally developed by Carath\'{e}odory to treat first
order regular systems. Generalizations to treat constrained systems were
developed by G\"{u}ler \cite{Guler 1}, \cite{Guler 2}, who dealed with
lagrangians of first order, and more recently by Pimentel and Teixeira
\cite{Rand Pim1}, \cite{Rand Pim2}, who worked with lagrangians of higher
order. An approach to treat Berezinian singular systems can also be found in
literature in a work of Pimentel, Teixeira and Tomazelli \cite{Rand Pim Jef}.
This formalism does not make use of the physical arguments, as it happens in
the hamiltonian one, but it uses the rigorous mathematical structure of the
partial differential equation theory, being this a significant feature of the
formalism. Although the development of Hamilton-Jacobi approach is quite
recent, it can be found in literature some important applications of this
formalism \cite{REF HJ}, \cite{REF HJ 1}, \cite{REF HJ2}, including an
application to Teleparallelism \cite{GRG 2003}.

The Teleparallel Equivalent of General Relativity (TEGR or Teleparallelism)
\cite{Teleparalelismo}, \cite{Hehl}, as the name suggests, is a theory of
gravitation that was built to be equivalent to General Relativity (GR). In
this way is expected that TEGR, as GR, be a singular system. Moreover, some
important aspects of Teleparallelism, as geodesics and ``force'' equations and
energy-momentum tensor, have been studied\cite{Ze Geraldo 0} (and references
cited therein), as well as the interactions of spin0, spin1 and spinor fields
in TEGR \cite{Ze Geraldo 1}, \cite{Ze Geraldo 2}, \cite{Ze Geraldo 3},
\cite{Tadeu}, \cite{Casana} and the gauge symmetries of this theory
\cite{Blago}. In \cite{GRG 2003} an application of Hamilton-Jacobi formalism
to TEGR was made using a gauge fixing and an ADM 3+1 decomposition of space
time \cite{ADM}.

In this work we intend to analyse the structure of Teleparallelism without
using an ADM decomposition and without fixing the time gauge condition, since
none of these conditions are requirements of the HJ approach. Accordingly,
first we will make a review of Hamilton-Jacobi approach to singular systems
and expose some important results of TEGR. Afterwards we analyse the
constraint structure of the theory, and we make some comments at last.

Here we use the following conventions: $\mu,\nu,...=0,i\,\left(
i=1,2,3\right)  $\thinspace are spacetime indices; $a,b,...=0,(i)$ $\left(
i=1,2,3\right)  $ are $SO\left(  3,1\right)  $ indices. The Minkowski metric
is fixed as $\left(  -+++\right)  .$ When the notation differs from this (as
it happens in the next section) we shall indicate explicitly.

\section{The Hamilton-Jacobi Formalism}

According to Carath\'{e}odory Equivalent Lagrangian method, given a Lagrangian
$L=L\left(  q^{i},\dot{q}^{i}\right)  $, $i=1,...,N$, another Lagrangian
$L^{\prime}=L^{\prime}\left(  q^{i},\dot{q}^{i}\right)  $,%
\begin{equation}
L^{\prime}\left(  q^{i},\dot{q}^{i}\right)  =L\left(  q^{i},\dot{q}%
^{i}\right)  -\frac{d}{dt}S\left(  q^{i},t\right)  , \label{L'}%
\end{equation}
can be found, in such a way that both action integrals, $A=\int dtL$ and
$A^{\prime}=\int dtL^{\prime}$, have simultaneous extremes\footnote{It is
important to notice that if $L$\ is a singular Lagrangian, i.e. $\det\left(
\frac{\partial^{2}L}{\partial\dot{q}^{i}\partial\dot{q}^{j}}\right)  =0$, then
$L^{\prime}$\ will also be singular because $\frac{\partial^{2}L^{\prime}%
}{\partial\dot{q}^{i}\partial\dot{q}^{j}}=\frac{\partial^{2}L}{\partial\dot
{q}^{i}\partial\dot{q}^{j}}$.}, $\delta A=\delta A^{\prime}$. Since the
variational problem of finding an extreme of the action integral is the same
for both Lagrangians,\ then $L$ and $L^{\prime}$ are said to be equivalent.

To find an extreme of the action $A^{\prime}$\ it is sufficient to find a set
of equations $\beta^{i}\left(  q^{j},t\right)  $\ such that
\begin{equation}
L^{\prime}\left(  q^{i},\dot{q}^{i}=\beta^{i}\left(  q^{j},t\right)  \right)
=L\left(  q^{i},\dot{q}^{i}\right)  -\frac{\partial}{\partial t}S\left(
q^{i},t\right)  -\frac{\partial S\left(  q^{i},t\right)  }{\partial q^{j}}%
\dot{q}^{j}=0, \label{L'=0}%
\end{equation}
and in a neighbourhood of $\dot{q}^{i}=\beta^{i}\left(  q^{j},t\right)  $\ the
condition $L^{\prime}\left(  q^{i},\dot{q}^{i}\right)  >0$ is satisified. From
the condition above it follows that%
\begin{equation}
p_{i}\equiv\left.  \frac{\partial L}{\partial\dot{q}^{i}}\right|  _{\dot
{q}^{i}=\beta^{i}}=\left.  \frac{\partial S}{\partial q^{i}}\right|  _{\dot
{q}^{i}=\beta^{i}}. \label{momenta}%
\end{equation}

For a singular Lagrangian $L$ the condition $\det H_{ij}=0$, where
$H_{ij}=\frac{\partial^{2}L}{\partial\dot{q}^{i}\partial\dot{q}^{j}}%
=\frac{\partial p_{j}}{\partial\dot{q}^{i}}$ is the Hessian matrix, is
satisfied. If the rank of this matrix is $P=N-R$, then the coordinates $q^{i}%
$\ may be ordered in such a way that the $P\mathsf{x}P$ matrix on the right
bottom corner of the Hessian matrix is nonsingular:%
\begin{equation}
\det H_{ab}=\det\frac{\partial p_{a}}{\partial\dot{q}^{b}}\neq
0,~\ \ \ \ \ \ a,b=R+1,...,N. \label{subHess}%
\end{equation}
This guarantees that $N-R$ velocities $\dot{q}^{b}$ can be written as $\dot
{q}^{b}=f^{b}\left(  q^{i},p_{a}\right)  $. The other momenta $p_{\alpha}%
$\ ($\alpha=1,...R$) do not depend on any velocity, but they may depend on the
coordinates, the time and the other momenta $p_{a}$, so that $R$ relations can
be written:%
\begin{equation}
p_{\alpha}=-H_{\alpha}\left(  t,q^{\beta}\equiv t^{\beta},q^{a},p_{a}\right)
. \label{vinc1}%
\end{equation}

It can also be seen that, using the standard definition $H_{0}\equiv p_{i}%
\dot{q}^{i}-L$, the expression (\ref{L'=0}) can be rewritten as\footnote{It
can be directly verified that $\frac{\partial H_{0}}{\partial\dot{q}^{\beta}%
}=0.$}%
\begin{equation}
p_{0}+H_{0}\left(  t,t^{\alpha},q^{a},p_{a}\right)  =0, \label{eqHJ}%
\end{equation}
where $p_{0}\equiv\frac{\partial S}{\partial t}$. The equations (\ref{vinc1})
and (\ref{eqHJ}) can be written in an unified way
\begin{equation}
H_{\alpha}^{\prime}\equiv p_{\alpha}+H_{\alpha}\left(  t^{\beta},q^{a}%
,p_{a}\right)  =0,~\ \ \ \ \alpha,\beta=0,1,...,R, \label{vinculos}%
\end{equation}
where $t^{0}\equiv t$\ \ This is a set of first order differential\ partial
equations, called \textit{Hamilton-Jacobi Partial Differential Equations} (HJPDE).

\subsection{Integrability Conditions}

It is known from the theory of partial differential equations \cite{Carat}
that a set of total differential equations, named as the characteristics
equations, is associated with a set of partial differential equations. It can
be verified, using the independence of the momenta and the fact that
$S=S\left(  t^{\alpha},q^{a}\right)  $ $\left(  \alpha
=0,1,...,R;~a=R+1,...,N\right)  $, that\ the characteristics equations of the
Hamilton-Jacobi equations are%
\begin{equation}
\left\{
\begin{array}
[c]{c}%
d\eta^{I}=E^{IJ}\frac{\partial H_{\alpha}^{\prime}}{\partial\eta^{J}%
}dt^{\alpha}=\left\{  \eta^{I},H_{\alpha}^{\prime}\right\}  dt^{\alpha},\\
dS=\frac{\partial S}{\partial q^{i}}dq^{i}=p_{i}\frac{\partial H_{\alpha
}^{\prime}}{\partial p_{i}}dt^{\alpha},
\end{array}
\right.  ~\ \
\begin{array}
[c]{c}%
I,J=\left(  \zeta;i\right)  ,\,\zeta=1,2;\\
\alpha=0,...,R;~i=0,...,N.
\end{array}
\label{eq carac}%
\end{equation}
where $\left\{  \eta^{1i}\right\}  =\left\{  q^{i}\right\}  $ and $\left\{
\eta^{2i}\right\}  =\left\{  p_{i}\right\}  $, $E^{IJ}=\delta_{\,j}^{i}\left[
\delta_{\,1}^{\zeta}\delta_{\,\sigma}^{2}-\delta_{\,2}^{\zeta}\delta
_{\,\sigma}^{1}\right]  ,$ $(I=\left(  \zeta,i\right)  ,\,J=\left(
\sigma,\,j\right)  ).$ We also have $\left\{  F,G\right\}  =\frac{\partial
F}{\partial\eta^{I}}E^{IJ}\frac{\partial G}{\partial\eta^{J}}$ , which is the
Poisson Brackets of $F$ and $G$. The equations of the first line of
(\ref{eq carac}) will be called from now on as equations of motion.

To assure that a unique solution to the Hamilton-Jacobi equations can be
found, it is enough to prove that the characteristics equations constitute a
set of integrable equations. In fact it is enough to prove that only the
equations of motion is a set of integrable equations, because if this
condition is satisfied, then the last of the characteristics equations will be
integrable as a consequence.

From the theory of differential equations it is also known that, associated
with a set of total equations, $dx^{I}=b_{\alpha}^{I}\left(  x^{J}\right)
dt^{\alpha}$, there are linear operators $X_{\alpha}$\ such that%
\begin{equation}
X_{\alpha}F\left(  x^{J}\right)  =b_{\alpha}^{I}\frac{\partial F}{\partial
x^{I}}=0. \label{eq dif parc}%
\end{equation}
As a consequence, for any function $F$\ at least twice differentiable, the
condition%
\begin{equation}
\left[  X_{\alpha},X_{\beta}\right]  F=\left(  X_{\alpha}X_{\beta}-X_{\beta
}X_{\alpha}\right)  F=0 \label{comut F}%
\end{equation}
will be satisfied. If the calculation of each quantity $\left[  X_{\alpha
},X_{\beta}\right]  F$\ is a linear combination of (\ref{eq dif parc}),
$\left[  X_{\alpha},X_{\beta}\right]  F=C_{\alpha\beta}^{~\ \gamma}X_{\gamma
}F,$ the partial differential equations $X_{\alpha}F=0$\ is said to be
\textit{complete}. Otherwise a new operator $X$\ can be defined, such that
$XF=0$, and it can be added to the initial set of operators $X_{\alpha}$. This
procedure can be repeated until a complete set of partial diferential be
obtained. The total differential equations will be integrable if the set of
associated partial equations is complete.

In the case of the equations of motion, we have $X_{\alpha}F=\left\{
F,H_{\alpha}^{\prime}\right\}  $, and $\left[  X_{\alpha},X_{\beta}\right]
F=-\left\{  F,\left\{  H_{\alpha}^{\prime},H_{\beta}^{\prime}\right\}
\right\}  $. Since the relation (\ref{comut F})\ must be satisfied for any
function $F$\ then it follows that%
\begin{equation}
\left\{  H_{\alpha}^{\prime},H_{\beta}^{\prime}\right\}  =0, \label{PP vinc=0}%
\end{equation}
or equivalently, considering the independence of $t^{\alpha}$,%
\begin{equation}
dH_{\alpha}^{\prime}=\left\{  H_{\alpha}^{\prime},H_{\beta}^{\prime}\right\}
dt^{\beta}=0. \label{cond int}%
\end{equation}

While these conditions are not satisfied, new relations of the type
$H^{\prime}=0$\ can be established, and the integrability conditions must be
tested until a complete set of partial differential equations is obtained.

\section{The Teleparallel Equivalent of General Relativity}

While General Relativity is a theory of gravitation built on Riemann%
\'{}%
s space-time, that is a particular case of Riemann-Cartan manifold with a nule
torsion tensor, Teleparallel Equivalent of General Relativity (TEGR) is a
theory of gravitation built on Weitzenb\"{o}ck space-time\cite{Weitzen}, which
is also a particular case of Riemann-Cartan manifold, where the torsion tensor
is non-vanishing and the curvature is nule. In this theory, usually studied in
its Lagrangian formulation, the dynamical fields are tetrads, $e_{\mu}^{a}$,
that constitute a set of four vectors under General Coordinate transformation
on space-time and four vectors under Lorentz transformation on tangent space.

To assure the vanishing of curvature of Weitzenb\"{o}ck space-time, we impose
the absolute parallelism condition, which implies that the spin connection is
null. With this condition, Cartan (or Weitzenb\"{o}ck) connection is
completely determined by tetrads, $\Gamma_{\mu\nu}^{\lambda}=e_{a}^{\lambda
}\partial_{\mu}e_{\nu}^{a}$, and the torsion, which is the antisymmetric part
of the connection, is given by $T_{\,\ \ \mu\nu}^{a}=\partial_{\mu}e_{\nu}%
^{a}-\partial_{\nu}e_{\mu}^{a}$. As it is well known, in TEGR the torsion
carries the information about the dynamics of the system. This can be directly
verified from the Lagrangian density,%
\begin{equation}
L=-ke\Sigma^{abc}T_{abc}\,\ \ , \label{L}%
\end{equation}
where%
\[
\Sigma^{abc}\equiv\frac{1}{4}\left(  T^{abc}+T^{bac}-T^{cab}\right)  +\frac
{1}{2}\left(  \eta^{ac}T^{b}-\eta^{ab}T^{c}\right)  \,\ \ ,
\]
$e\equiv\det\left(  e_{\mu}^{a}\right)  $, $T^{b}\equiv\eta_{ab}T^{abc}$, and
$k=1/(16\pi G)$ ($G$ is the gravitacional constant). The equations of motion
of the theory (in the absence of matter's field) can be obtained from $L$ with
the use of a variational principle:%
\begin{equation}
\partial_{\mu}\left(  e\Sigma_{g}^{\,\mu\alpha}\right)  +\Sigma_{a}%
^{\,\alpha\mu}T_{\,\ g\mu}^{a}-e\frac{1}{4}e_{g}^{\,\alpha}\Sigma^{abc}%
T_{abc}=0\,\ \ . \label{Eq motion Lagr}%
\end{equation}
It is important to remind that this Lagrangian density is built to be
equivalent, up to a surface term, to Einstein-Hilbert Lagrangian density of
General Relativity \cite{Blagojevic}.

In order to study the constraints of this theory using Hamilton-Jacobi
formalism, we must know the structure of TEGR in phase space. Following the
steps of Carath\'{e}odory Lagrangian Equivalent method, first we must choose
some coordinate to perform the role of time, $t$. The choice employed here is
the same that it was used by Dirac \cite{Dirac1}, \cite{Dirac2} in his study
of General Relativity and consists of taking the zeroth coordinate as $t$.

Secondly we must obtain the momenta canonically conjugated to $e_{a\mu}$. This
can be done with the definition%
\begin{equation}
\Pi^{a\mu}\equiv\frac{\delta L}{\delta\dot{e}_{a\mu}}\,\ \ , \label{Pi^a_mi}%
\end{equation}
from where it follows
\begin{align}
\Pi^{a\mu}  &  =-4ke\Sigma^{a0\mu}\Rightarrow\nonumber\\
&  \Rightarrow\left\{
\begin{array}
[c]{l}%
\Pi^{ak}=ke\left[  g^{00}\left(  -g^{kj}T_{\,0j}^{a}-e^{aj}T_{\,0j}%
^{k}+2e^{ak}T_{\,0j}^{j}\right)  \right.  +\\
+g^{0k}\left(  g^{0j}T_{\,0j}^{a}+e^{aj}T_{\,0j}^{0}\right)  +e^{a0}\left(
g^{0j}T_{\,0j}^{k}+g^{kj}T_{\,0j}^{0}\right)  +\\
-2\left(  e^{a0}g^{0k}T_{\,0j}^{j}+e^{ak}g^{0j}T_{\,0j}^{0}\right)  +\\
-g^{0i}g^{kj}T_{\,ij}^{a}+e^{ai}\left(  g^{0j}T_{\,ij}^{k}-g^{kj}T_{\,ij}%
^{0}\right)  +\\
\left.  -2\left(  g^{0i}e^{ak}-g^{ik}e^{a0}\right)  T_{\,ji}^{j}\right]
\,\ \ ,\\
\Pi^{a0}=0\,\ \ .
\end{array}
\right.  . \label{Pi^a_0 =0}%
\end{align}

This last result arises from the fact that there is no ``time'' derivative of
$e_{a0}$ in $L$, and as a consequence we have four constraints,%
\begin{equation}
H_{1}^{\prime a}=\int d^{3}y\Pi^{a0}\left(  y\right)  =0 \label{H'_1}%
\end{equation}

The next step is to obtain the Hamiltonian density and this may be achieved
with the employment of the prescription $``L=p\dot{q}-H_{0}"$, and, obviously,
with the definitions (\ref{Pi^a_mi}):%
\begin{align}
H_{0}  &  =-e_{a0}\partial_{k}\Pi^{ak}-\frac{1}{4g^{00}}ke\left(  g_{ik}%
g_{jl}P^{ij}P^{kl}-\frac{1}{2}P^{2}\right)  +\nonumber\\
&  +ke\left(  \frac{1}{4}g^{im}g^{nj}T_{\,mn}^{a}T_{aij}+\frac{1}{2}%
g^{nj}T_{\,mn}^{i}T_{\,ij}^{m}-g^{ij}T_{\,ji}^{j}T_{\,nk}^{n}\right)  \,\ \ ,
\label{H_0}%
\end{align}
where%
\begin{align*}
P^{ik}  &  \equiv\frac{1}{ke}\Pi^{\left(  ik\right)  }-\Delta^{ik}\,\ \ ,\\
P  &  \equiv g_{ik}P^{ik}\,\ \ ,
\end{align*}
with%
\[
\Delta^{ik}\equiv-g^{0m}\left(  g^{kj}T_{\,mj}^{i}+g^{ij}T_{\,mj}^{k}%
-2g^{ik}T_{\,mj}^{j}\right)  -\left(  g^{km}g^{0i}+g^{im}g^{0k}\right)
T_{\,mj}^{j}\,\ \ ,
\]
and $\Pi^{\left(  ik\right)  }$ being the symmetric part of $\Pi^{ik}\equiv
e_{a}^{\,i}\Pi^{ak}$.

As it was seen in the previous section, the canonical Hamiltonian, that is
obtained from the spatial integration of the Hamiltonian density, satisfies
the Hamilton-Jacobi equation that may be rewritten as%
\begin{equation}
H_{0}^{\prime}=\int d^{3}y\left(  P_{0}\left(  y\right)  +H_{0}\left(
y\right)  \right)  =0\,\ \ , \label{H'_0}%
\end{equation}
if we define a density of momenta canonically conjugated to time, $P_{0}$.

The equations (\ref{H'_1}) and (\ref{H'_0}) constitute the set of
Hamilton-Jacobi Partial Differential Equations (HJPDE) of TEGR. With these
results we can write the equations of motion as%
\begin{equation}
d\eta^{I}=\left\{  \eta^{I},H_{0}^{\prime}\right\}  dt+\left\{  \eta^{I}%
,H_{1}^{\prime a}\right\}  de_{a0}~\ \ , \label{dXi}%
\end{equation}
where $\eta^{I}=\left\{  e_{a\mu},t;\Pi^{a\mu},P_{0}\right\}  $, and we shall
verify if they constitute a integrable system.

\subsection{Integrability Conditions}

The integrability conditions of the system of total differential equations
(\ref{dXi}) are given by%
\begin{align*}
dH_{0}^{\prime}  &  =\left\{  H_{0}^{\prime},H_{0}^{\prime}\right\}
dt+\left\{  H_{0}^{\prime},H_{1}^{\prime a}\right\}  de_{a0}~\ \ ,\\
dH_{1}^{\prime b}  &  =\left\{  H_{1}^{\prime b},H_{0}^{\prime}\right\}
dt+\left\{  H_{1}^{\prime b},H_{1}^{\prime a}\right\}  de_{a0}~\ \ .
\end{align*}

These Poisson Brackets can be evaluated if we notice that in this theory there
is no explicit dependence on time which implies
\[
\left\{  F,P_{0}\right\}  =0~\ \ ,
\]
and that%
\[
\left\{  F,H_{1}^{\prime a}\right\}  =\int d^{3}x\frac{\delta F}{\delta
e_{a0}\left(  x\right)  }\,\ \ .
\]

Both expressions are valid for an arbitrary $F$. With these results we see
that%
\begin{align}
\left\{  H_{0}^{\prime},H_{0}^{\prime}\right\}   &  =\int d^{3}yd^{3}z\left\{
H_{0}\left(  y\right)  ,H_{0}\left(  z\right)  \right\}  =-\int d^{3}%
yd^{3}z\left\{  H_{0}\left(  z\right)  ,H_{0}\left(  y\right)  \right\}
=0\,\ \ ,\label{PP_H0_H0}\\
\left\{  H_{1}^{\prime b},H_{1}^{\prime a}\right\}   &  =\int d^{3}%
yd^{3}z\frac{\delta\Pi^{b0}\left(  z\right)  }{\delta e_{a0}\left(  y\right)
}=0\,\ \ ,\label{PP_H1_H1}\\
\left\{  H_{0}^{\prime},H_{1}^{\prime a}\right\}   &  =\int d^{3}yC^{a}\left(
y\right)  \,\ \ , \label{PP_H0_H1}%
\end{align}
where%
\[
C^{a}\equiv\frac{\delta H_{0}}{\delta e_{a0}}=e^{a0}H_{0}+e^{ai}F_{i}\,\ \ ,
\]
with%
\begin{align*}
F_{i}  &  \equiv H_{i}+\Gamma^{m}T_{0mi}+\Gamma^{km}T_{kmi}+\frac{1}{2g^{00}%
}\left(  g_{ik}g_{jl}P^{kl}-\frac{1}{2}g_{ij}P\right)  \Gamma^{j}~\ \ ,\\
H_{i}  &  \equiv-e_{bi}\partial_{k}\Pi^{bk}-\Pi^{bk}T_{bki}~\ \ ,\\
\Gamma^{k}  &  \equiv\Pi^{0k}+2ke\left(  g^{kj}g^{0i}T_{\,ij}^{0}-g^{0k}%
g^{0i}T_{\,ji}^{j}+g^{00}g^{ik}T_{\,ij}^{j}\right)  ~\ \ ,\\
\Gamma^{ik}  &  \equiv-\Gamma^{ki}=\Pi^{\left[  ik\right]  }+ke\left[
-g^{im}g^{kj}T_{\,mj}^{0}+\left(  g^{im}g^{0k}-g^{km}g^{0i}\right)
T_{\,mj}^{j}\right]  ~\ \ .
\end{align*}
This decomposition of $C^{a}$\ is very important at this stage \cite{Maluf &
RN}\ and showed to be essential for the further steps.

If we want the system of total differential equations to be integrable, the
condition%
\begin{equation}
C^{a}=0 \label{C^a=0}%
\end{equation}
must be satisfied. Moreover, if we recall the orthogonality of the tetrads, we
have:%
\begin{align*}
H_{0}  &  =e_{a}^{~0}C^{a}=0~\ \ ,\\
F^{i}  &  =e_{a}^{~i}C^{a}=0~\ \ .
\end{align*}

The condition $C^{a}=0$ lead us to define four new constraints,%
\begin{equation}
H_{2}^{\prime a}=\int d^{3}yC^{a}\left(  y\right)  =0~\ \ . \label{Vinc Sec1}%
\end{equation}
These constraints must be incorporated to the previous set of constraints and
the integrability conditions must be tested again:%
\begin{align*}
dH_{0}^{\prime}  &  =\left\{  H_{0}^{\prime},H_{0}^{\prime}\right\}
dt+\left\{  H_{0}^{\prime},H_{1}^{\prime a}\right\}  de_{a0}+\left\{
H_{0}^{\prime},H_{2}^{\prime a}\right\}  d\lambda_{a}~\ \ ,\\
dH_{1}^{\prime b}  &  =\left\{  H_{1}^{\prime b},H_{0}^{\prime}\right\}
dt+\left\{  H_{1}^{\prime b},H_{1}^{\prime a}\right\}  de_{a0}+\left\{
H_{1}^{\prime b},H_{2}^{\prime a}\right\}  d\lambda_{a}~\ \ ,\\
dH_{2}^{\prime b}  &  =\left\{  H_{2}^{\prime b},H_{0}^{\prime}\right\}
dt+\left\{  H_{2}^{\prime b},H_{1}^{\prime a}\right\}  de_{a0}+\left\{
H_{2}^{\prime b},H_{2}^{\prime a}\right\}  d\lambda_{a}~\ \ ,
\end{align*}
where $\lambda_{a}$ are four parameters that could not be associated with
specific fields of the theory, because the constraint $H_{2}^{\prime a}$ is
not a constraint of the type $``H+p=0"$. We can evaluate those Poisson
Brackets one by one:%

\[
\left\{  H_{0}^{\prime},H_{2}^{\prime a}\right\}  =\int d^{3}yd^{3}z\left\{
H_{0}\left(  y\right)  ,C^{a}\left(  z\right)  \right\}  ~\ \ ,
\]%
\[
\left\{  H_{1}^{\prime b},H_{2}^{\prime a}\right\}  =\int d^{3}yd^{3}z\left\{
\Pi^{a0}\left(  y\right)  ,C^{a}\left(  z\right)  \right\}  =\int d^{3}%
yd^{3}z\frac{\delta C^{b}\left(  z\right)  }{\delta e_{a0}\left(  y\right)
}~\ \ ,
\]%
\[
\left\{  H_{2}^{\prime b},H_{2}^{\prime a}\right\}  =\int d^{3}yd^{3}z\left\{
C^{b}\left(  y\right)  ,C^{a}\left(  z\right)  \right\}  ~\ \ .
\]

It can be verified that
\begin{equation}
\frac{\delta\Gamma^{ik}}{\delta e_{a0}}=-\frac{1}{2}\left(  e^{ai}\Gamma
^{k}-e^{ak}\Gamma^{i}\right)  ~\ \ , \label{delta gama_ik delta e_a0}%
\end{equation}%
\begin{equation}
\frac{\delta\Gamma^{i}}{\delta e_{a0}}=-e^{a0}\Gamma^{k}~\ \ .
\label{delta gama_ik_delta e_a0}%
\end{equation}
These results lead us to conclude that%
\[
\left\{  H_{1}^{\prime b},H_{2}^{\prime a}\right\}  =-\int d^{3}ye^{ai}%
e^{bn}\left(  \frac{1}{2k}g_{in}g_{jl}\Gamma^{l}+\left(  T_{inj}%
+T_{jni}\right)  \right)  \Gamma^{j}~\ \ ,
\]
and if we wish that these relations were nule for any $e^{a\mu}$, then%
\begin{equation}
\Gamma^{j}=0~\ \ . \label{gama_j=0}%
\end{equation}

We also have the following result%
\[
\left\{  H_{0}^{\prime},H_{2}^{\prime a}\right\}  =\int d^{3}yd^{3}z\left\{
H_{0}\left(  y\right)  ,C^{a}\left(  z\right)  \right\}  =\int d^{3}%
yd^{3}ze^{ai}\left\{  H_{0}\left(  y\right)  ,F_{i}\left(  z\right)  \right\}
,
\]
which may be evaluated, considering that $\Gamma^{j}=0$,\ by the following
Poisson Brackets:%
\begin{equation}
\left\{  H_{0}\left(  y\right)  ,H_{i}\left(  z\right)  \right\}
=H_{0}\left(  z\right)  \frac{\partial}{\partial y^{i}}\delta\left(
y-z\right)  -C^{a}\partial_{i}e_{a0}\delta\left(  y-z\right)  ~\ \ ,
\label{PP_H0_Hi}%
\end{equation}%
\begin{align}
\left\{  H_{0}\left(  y\right)  ,\Gamma^{i}\left(  z\right)  \right\}   &
=\left[  g^{0i}H_{0}-\frac{1}{g^{00}}P^{kl}\left(  g_{kj}g_{ml}-\frac{1}%
{2}g_{kl}g_{jm}\right)  g^{0j}\Gamma^{mi}+\right. \nonumber\\
&  +\left(  \Gamma^{ni}e^{a0}+\Gamma^{n}e^{ai}\right)  \partial_{n}%
e_{a0}+\frac{1}{2}\Gamma^{mn}T_{~nm}^{i}+\nonumber\\
&  \left.  +2\partial_{n}\Gamma^{ni}+g^{in}\left(  H_{n}-\Gamma^{j}%
T_{0nj}-\Gamma^{mj}T_{mnj}\right)  \right]  \delta\left(  y-z\right)
+\nonumber\\
&  +\Gamma^{ni}\left(  y\right)  \frac{\partial}{\partial y^{n}}\delta\left(
y-z\right)  ~\ \ , \label{PP_H0_gama_i}%
\end{align}%
\begin{align}
\left\{  H_{0}\left(  y\right)  ,\Gamma^{ij}\left(  z\right)  \right\}   &
=\left[  -\frac{1}{2g^{00}}P^{kl}\left(  g_{km}g_{nl}-\frac{1}{2}g_{kl}%
g_{mn}\right)  \left(  g^{mi}\Gamma^{nj}-g^{mj}\Gamma^{ni}\right)  +\right.
\nonumber\\
&  \left.  +\frac{1}{2}\left(  \Gamma^{nj}e^{ai}-\Gamma^{ni}e^{aj}\right)
\partial_{n}e_{a0}\right]  \delta\left(  y-z\right)  ~\ \ .
\label{PP_H0_gama_ij}%
\end{align}
Again, if we want $\left\{  H_{0}^{\prime},H_{2}^{\prime a}\right\}  $\ to
vanish for any tetrad, we must take%
\begin{equation}
\Gamma^{ij}=0~\ \ . \label{gama_ij=0}%
\end{equation}

The expressions (\ref{gama_j=0}) and (\ref{gama_ij=0}) make us define new
constraints%
\[
H_{3}^{\prime i}=\int d^{3}y\Gamma^{i}\left(  y\right)  =0~\ \ ,
\]%
\[
H_{4}^{\prime ij}=\int d^{3}y\Gamma^{ij}\left(  y\right)  =0~\ \ ,
\]
that we must add to the previous set of constraints
\begin{align*}
dH_{0}^{\prime}  &  =\left\{  H_{0}^{\prime},H_{0}^{\prime}\right\}
dt+\left\{  H_{0}^{\prime},H_{1}^{\prime a}\right\}  de_{a0}+\left\{
H_{0}^{\prime},H_{2}^{\prime a}\right\}  d\lambda_{a}+\\
&  +\left\{  H_{0}^{\prime},H_{3}^{\prime i}\right\}  d\omega_{i}+\left\{
H_{0}^{\prime},H_{4}^{\prime ij}\right\}  d\omega_{ij}~\ \ ,\\
&  \vdots\\
dH_{4}^{\prime mn}  &  =\left\{  H_{4}^{\prime mn},H_{0}^{\prime}\right\}
dt+\left\{  H_{4}^{\prime mn},H_{1}^{\prime a}\right\}  de_{a0}+\left\{
H_{4}^{\prime mn},H_{2}^{\prime a}\right\}  d\lambda_{a}+\\
&  +\left\{  H_{4}^{\prime mn},H_{3}^{\prime i}\right\}  d\omega_{i}+\left\{
H_{4}^{\prime mn},H_{4}^{\prime ij}\right\}  d\omega_{ij}~\ \ ,
\end{align*}
and once more test the integrability conditions. We must observe that
$\omega_{i}$ and $\omega_{ij}$\ are, as $\lambda_{a}$, parameters in the theory.

By direct inspection one can verify that these Poisson Brackets constitute an
algebra, what may be verified with the above results and the following ones:%
\begin{equation}
\left\{  H_{j}\left(  y\right)  ,H_{k}\left(  z\right)  \right\}
=-H_{k}\left(  y\right)  \frac{\partial}{\partial y^{j}}\delta\left(
y-z\right)  -H_{j}\left(  y\right)  \frac{\partial}{\partial y^{k}}%
\delta\left(  y-z\right)  ~\ \ , \label{PP_Hj_Hk}%
\end{equation}%
\begin{equation}
\left\{  \Gamma^{i}\left(  y\right)  ,\Gamma^{j}\left(  z\right)  \right\}
=0~\ \ , \label{PP_gama_i_gama_j}%
\end{equation}%
\begin{equation}
\left\{  \Gamma^{ij}\left(  y\right)  ,\Gamma^{kl}\left(  z\right)  \right\}
=\frac{1}{2}\left(  g^{il}\Gamma^{jk}+g^{jk}\Gamma^{il}-g^{ik}\Gamma
^{jl}-g^{jl}\Gamma^{ik}\right)  \delta\left(  y-z\right)  ~\ \ ,
\label{PP_gama_ij_gama_kl}%
\end{equation}%
\begin{equation}
\left\{  \Gamma^{ij}\left(  y\right)  ,\Gamma^{k}\left(  z\right)  \right\}
=\left(  g^{oj}\Gamma^{ki}-g^{0i}\Gamma^{kj}\right)  \delta\left(  y-z\right)
~\ \ , \label{PP_gama_ij_gama_k}%
\end{equation}%
\begin{align}
\left\{  H_{i}\left(  y\right)  ,\Gamma^{j}\left(  z\right)  \right\}   &
=\delta_{i}^{j}\Gamma^{n}\left(  z\right)  \frac{\partial}{\partial z^{k}%
}\delta\left(  y-z\right)  +\Gamma^{j}\left(  y\right)  \frac{\partial
}{\partial y^{i}}\delta\left(  y-z\right) \nonumber\\
&  -\Gamma^{j}e^{a0}\partial_{i}e_{a0}\delta\left(  y-z\right)  ~\ \ ,
\label{PP_Hi_gama_j}%
\end{align}%
\begin{align}
\left\{  H_{k}\left(  y\right)  ,\Gamma^{ij}\left(  z\right)  \right\}   &
=\Gamma^{ij}\left(  y\right)  \frac{\partial}{\partial y^{k}}\delta\left(
y-z\right)  +\nonumber\\
&  +\left(  \delta_{k}^{j}\Gamma^{ni}\left(  z\right)  -\delta_{k}^{i}%
\Gamma^{nj}\left(  z\right)  \right)  \frac{\partial}{\partial y^{k}}%
\delta\left(  y-z\right)  +\nonumber\\
&  +\frac{1}{2}\left(  e^{aj}\Gamma^{i}-e^{ai}\Gamma^{j}\right)  \partial
_{k}e_{a0}\delta\left(  y-z\right)  ~\ \ . \label{PP_Hk_gama_ij}%
\end{align}
As a consequence, no new constraint arises and then we conclude that the
system of total differential equations of motion, that now read%
\begin{align}
d\eta^{I}  &  =\left\{  \eta^{I},H_{0}^{\prime}\right\}  dt+\left\{  \eta
^{I},H_{1}^{\prime a}\right\}  de_{a0}+\left\{  \eta^{I},H_{2}^{\prime
a}\right\}  d\lambda_{a}+\nonumber\\
&  +\left\{  \eta^{I},H_{3}^{\prime i}\right\}  d\omega_{i}+\left\{  \eta
^{I},H_{4}^{\prime ik}\right\}  d\omega_{ik}~\ \ , \label{eq motion}%
\end{align}
is an integrable system.

\section{Concluding Remarks}

The analysis of the constraints of TEGR \textit{via} Hamilton-Jacobi shows
that the total differential equations of motions constitute a set of
integrable system, which implies that the HJPDE compose a system of partial
differential equations that has precisely a solution. Moreover, if we analyse
the set of equations (\ref{eq motion}) we see that phase space of the theory
is a subspace of the space $\left\{  e_{a\mu},t;\Pi^{a\mu},P_{0}\right\}  $.
We can affirm this because, according to Hamilton-Jacobi approach, the
variables $t$, $e_{a0}$, $\lambda_{a}$, $\omega_{i}$, $\omega_{ik}$ have the
status of parameters.

We can likewise compare the results obtained in this work with those obtained
by Maluf and Rocha-Neto \cite{Maluf & RN}, who studied the constraints of TEGR
with Dirac approach. We see that the constraints obtained here are exactly the
same obtained in \cite{Maluf & RN} (compare equations (\ref{H_0}),
(\ref{C^a=0}), (\ref{gama_j=0}), (\ref{gama_ij=0}) obtained here with
equations (18), (22), (20), (19) of \cite{Maluf & RN}) and consequently the
constraint algebra of the last is also the same algebra of this work. This is
a consequence of the equivalency between HJ integrability conditions and Dirac
consistency conditions (see the appendix in \cite{Rand Pim Jef}). However we
must notice that the paths that lead to these results in one and another
approach are very different \ One of the main differences that we point out is
that in Dirac approach we need to determine the set of primary constraints to
test the consistency conditions in order to obtain the secondary constraints.
This procedure was used in \cite{Maluf & RN}. In HJ we do not need to
distinguish between primary and secondary constraints. This characteristic
allowed us to work with an incomplete set of primary constraints (in Dirac's
language), and, with the employment of the integrability conditions, we were
able to find the secondary constraints and all the missing primary ones. This
is a peculiar characteristic of HJ approach.

Moreover we believe that it is possible to write the equations of motion
(\ref{eq motion}) for the dynamical variables in a form independent of the
parameters $\lambda_{a}$, $\omega_{i}$, $\omega_{ik}$, which would reveal the
gauge independence of the system's evolution, as it happens, for example, in
the study of QED \cite{Rand Pim Jef}.

\bigskip

{\Large Acknowledgements:}

B. M.\ Pimentel thanks CNPq and Funda\c{c}\~{a}o de Amparo \`{a} Pesquisa do
Estado de S\~{a}o Paulo, FAPESP,\ (grant number 02/00222-9) for partial
support; P. J. Pompeia thanks the staff of CTA for incentive and support; J.
F. da Rocha-Neto would like to thank Professor J. Geraldo Pereira for his
hospitality at the Instituto de F\'{\i}sica Te\'{o}rica IFT/UNESP and FAPESP
(grant number 01/00890-9) for partial support.

\end{document}